\documentclass[journal,times,onecolumn,12pt]{IEEEtran}
\usepackage{cite}

\usepackage[margin=1in]{geometry}
\usepackage{graphicx} 
\usepackage{caption}
\usepackage{subcaption}
\usepackage{float}
\usepackage{subfloat}
\graphicspath{{../Figures/}}
\usepackage{epsfig}
\usepackage{epstopdf}

\usepackage[american]{circuitikz}
\usepackage{tikz}
\usetikzlibrary{arrows}

\usepackage{siunitx}
\usepackage{booktabs} 
\usepackage{multirow}

\usepackage[cmex10]{amsmath}

\usepackage{array}

\begin{document}

\title{Performance Analysis of SSHC Rectifiers used in Ultrasonic Wireless Power Transfer}

\author{\IEEEauthorblockN{Sijun Du\IEEEauthorrefmark{1}}

\IEEEauthorblockA{\small \IEEEauthorrefmark{1}Department of Electrical Engineering and Computer Sciences, University of California at Berkeley, Berkeley, CA, 94720, USA.
}\vspace*{-3em}
}

\maketitle


\begin{abstract}

One of the key design considerations for biomedical implants is to minimize the system volume while achieving high performance. An attractive approach to power biomedical implants is to use wireless power transfer (WPT) with ultrasound. To receive the ultrasonic energy, a piezoelectric transducer is implemented, which is designed to have the resonance frequency at the frequency of the ultrasonic wave. To extract the received energy from the transducer, an energy rectification circuit is typically required. The commonly used rectifier is called a synchronized switch harvesting on inductor (SSHI) rectifier, which employs an inductor to increase the energy efficiency. However, the employment of inductor limits the system miniaturization since it usually needs to be around 1's mH. The synchronized switch harvesting on capacitors (SSHC) rectifier was recently proposed to be used on vibration energy harvesting and this architecture achieves inductor-less fully integrated design. In a vibration energy harvesting system, the piezoelectric transducer usually has large inherent capacitance and low resonant frequency. But the piezoelectric transducer used in ultrasonic WPT typically has very small capacitance and much higher resonant frequency. In this paper, the performance of a SSHC rectifier used in ultrasonic WPT is analyzed. The analysis covers system size and performance. Based on the analysis, this paper presents an important design consideration on designing a SSHC rectifier to be used in ultrasonic WPT, which is the allowed maximal ON resistance in the capacitor-to-capacitor charging loop in function of the transducer capacitance, transducer resonant frequency and the number of SSHC stages. 

\end{abstract}

\section{Introduction} \label{1909_intro}

In the past decade, piezoelectric energy harvesting (PEH) has drawn much research interests in various applications, such as autonomous wireless sensors in Internet of Things (IoT) networks, wearable electronics and biomedical implantable devices \cite{Mitcheson2008}. To be employed in these systems, the key design considerations for a PEH system are energy efficiency and system miniaturization. For miniaturization reasons, MEMS process is widely used to fabricate PEHs, as well as different kinds of sensors, such as inertial sensors, accelerometers and gyroscopes. A commonly used rectification circuit is the full-bridge rectifier (FBR) due to its simplicity and stability. Besides, this purely passive rectifier does not require any $V_{DD}$ to operate; hence, it eliminates the requirement of system cold-startup ability. However, the FBR usually sets high voltage threshold, which significantly reduces the overall energy efficiency \cite{Blystad2010pmehs}. In order to address this issue, the Synchronized Switch Harvesting on Inductor (SSHI) rectifier has been introduced by synchronously flipping the voltage across the piezoelectric transducer (PT). Many recent works have shown the huge performance enhancement due to the SSHI architecture \cite{badel2005efficiency, Szarka2013upfabrfe, Aktakka2014}. 

Despite the performance enhancement of the SSHI rectifier, it employs an inductor to perform energy conversion. In order to achieve good performance, the RLC oscillation loop should have a large inductance and a very small DC resistance. Typically, the inductor used in an SSHI rectifier is between 1 mH and 10 mH, which is relatively huge compared to other components in the energy harvesting system. To achieve system miniaturization, the inductor should be removed without sacrificing the performance. 

\section{SSHC interface circuit} \label{1909_sshcic}

\begin{figure}
\centering
\includegraphics[width=0.95\linewidth]{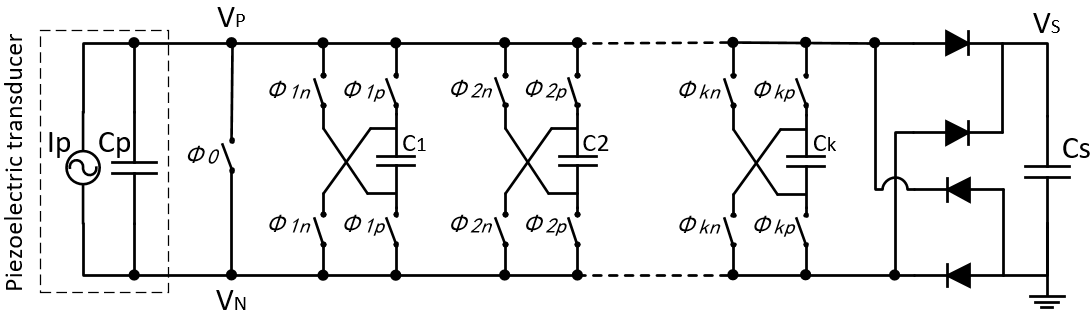}
\caption{Architecture of the SSHC rectifier. }
\label{fig:1909_hsscwkcaps}
\end{figure}

Fig. \ref{fig:1909_hsscwkcaps} shows the architecture of the SSHC rectifier, which employs a number of capacitors to flip the voltage across the transducer, instead of using an inductor. The number of capacitors, $k$, can be any positive integer. For a given $k$, there are $2k+1$ control signal phases to control the $4k+1$ switches. When the voltage across the piezoelectric transducer (PT) needs to be flipped from positive to negative, the $2k+1$ signals are sequenced as $\phi_{1p} \rightarrow \phi_{2p} \rightarrow \dots \rightarrow \phi_{kp-1} \rightarrow \phi_{kp} \rightarrow \phi_0 \rightarrow \phi_{kn} \rightarrow \phi_{kn-1} \rightarrow \dots \rightarrow \phi_{2n} \rightarrow \phi_{1n}$. The first $k$ phases are to dump the charge from the PT to the $k$ capacitors sequentially in the same polarization. The middle phase $\phi_0$ clears the remaining charge in the PT. The last $k$ phases are to transfer charge from the $k$ capacitor back to the PT in a reversed order and in an opposite polarization. 

\begin{figure}
\centering
\includegraphics[width=0.45\linewidth]{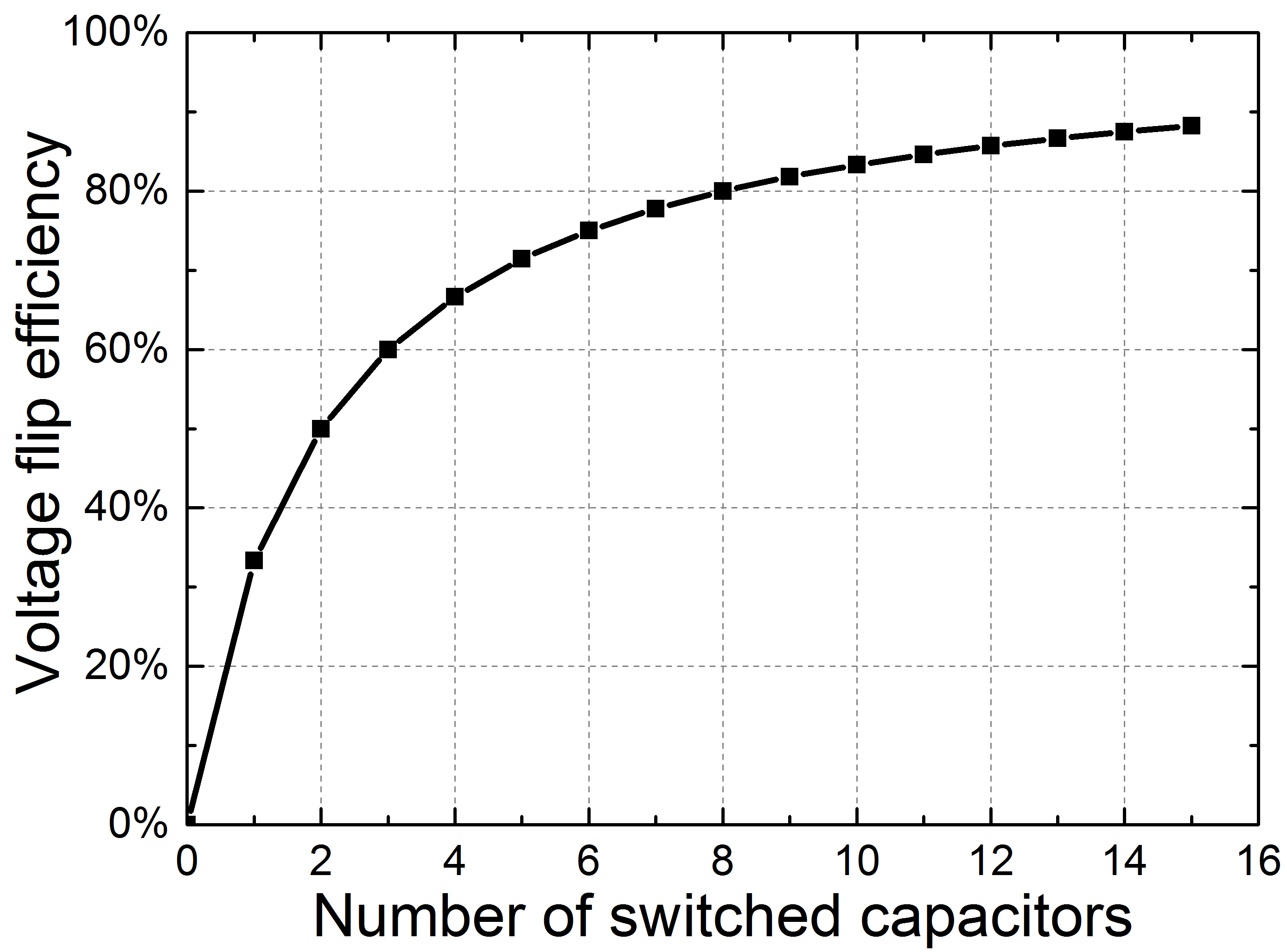}
\caption{Optimal voltage flip efficiency along different numbers of employed capacitors. }
\label{fig:1909_thevfecalk}
\end{figure}

When all the capacitors have equal capacitance and equal to the inherent capacitance of the PT, $C_P$, the voltage flip efficiency shown in Fig. \ref{fig:1909_thevfecalk} is achieved according to different numbers of employed capacitors. The figure shows that for a 1-stage SSHC rectifier (with 1 capacitor), the flip efficiency can achieve 33\%. This efficiency can be as high as 80\% when 8 capacitors are employed. When more capacitors are employed, the efficiency will increase further. 

The capacitor $C_P$ is the inherent capacitor of the PT, which is usually a few nF to 10's nF for a vibration energy harvester targeting at 100's Hz resonance frequency. In order to achieve the performance illustrated in Fig. \ref{fig:1909_thevfecalk}, all the $C_k$ capacitors need to be equal to $C_P$; hence they cannot be implemented on-chip since the total capacitance for a 8-stage SSHC rectifier will easily exceed 100 nF, which is very impractical to be implemented on-chip. A split-electrode design has been proposed recently to significantly reduce the required capacitance to achieve good performance, which enables full integration of all the capacitors. However, it requires the thick oxide process of the switch transistors to support high gate voltage levels, which is not preferred in standard CMOS processes. 

\section{SSHC in ultrasonic wireless power transfer} \label{1909_siuwpt}

Although the conventional SSHC rectifier without the split-electrode design cannot realize fully integrated capacitors for PTs for vibration energy harvesting, it can be used in PTs with very small inherent capacitors, such as PTs used in ultrasonic wireless power transfer (WPT). A PT to be employed as an energy receiver of ultrasonic WPT is typically with the size smaller than \SI{1}{mm^3}. The resonant frequency is typically around \SI{100}{kHz} with inherent capacitance smaller than 100 pF. When employing an 8-stage SSHC rectifier, each of the switched capacitors need to be equal to the capacitance of the PT, $C_P$. Assuming $C_P = \SI{100}{pF}$, each switched capacitor needs to have the capacitance equal to \SI{100}{pF}. Hence, the required total on-chip capacitance of a 8-stage SSHC rectifier is only around \SI{800}{pF}. This capacitance level is easy to be implemented on-chip with many CMOS processes using metal-insulator-metal (MIM) capacitors. In this section, the performance of a ultrasonic wireless power transfer (WPT) system using SSHC rectifiers will be analyzed on various aspects. 

\subsection{System size} \label{1909_syssize}

In a typical \SI{0.18}{\micro m} CMOS process, the metal-insulator-metal (MIM) capacitor is with a density of around \SI{2}{fF/\micro m^2}. This may differ for different processes. Assuming the inherent capacitor of the PT used in ultrasonic WPT is \SI{100}{pF}, each on-chip switched capacitor employed in the SSHC rectifier should also be \SI{100}{pF}, which takes around \SI{0.05}{mm^2} for a typical \SI{0.18}{\micro m} CMOS process. When employing a 8-stage SSHC rectifier, the total required on-chip capacitors occupy a chip area of around \SI{0.4}{mm^2}. The additional system size due to the switched capacitors should be less than \SI{1}{mm^3}. However, if an SSHI rectifier is employed, a \SI{5.6}{mH} inductor is typically required to achieve the 80\% voltage flip efficiency, equal to the performance of a SSHC rectifier employing 8 capacitors. For a \SI{5.6}{mH} inductor with several ohms of DC resistance, the size is typically around or larger than \SI{1}{cm^3}, which is 1000 times larger than that of the aforementioned 8 on-chip capacitors. From this perspective, conventional SSHC rectifiers can be easily employed into ultrasonic WPT systems without splitting the electrode of the PT. Due to the small $C_P$ of the ultrasonic PT, the on-chip capacitors occupy very small chip area.

\subsection{Output power analysis} \label{1909_vfdae}

\begin{figure}
\centering
\includegraphics[width=0.5\linewidth]{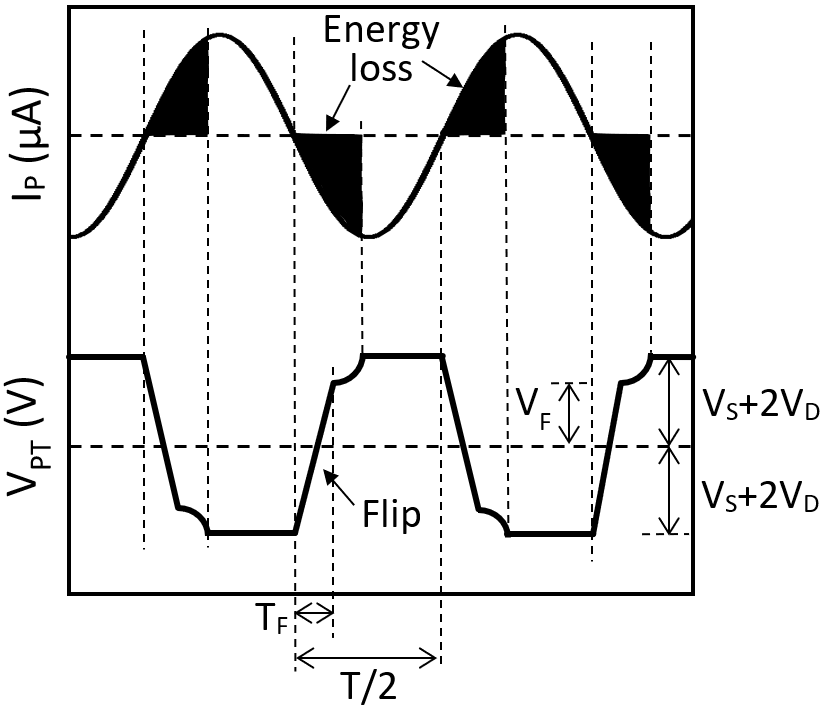}
\caption{Waveforms of $I_P$ and $V_{PT}$ of an SSHC rectifier on a Ultrasonic PT. }
\label{fig:1909_sshcwf}
\end{figure}

Although ultrasonic PT has very small inherent capacitance, $C_P$, which enables full integration of all the switched capacitors, the high resonant frequency of an ultrasonic PT may limit the actual performance of an SSHC. As aforementioned, the resonant frequency of a sub-mm sized ultrasonic PT is typically around \SI{100}{kHz}. Hence, the period of a such PT is \SI{10}{\micro s} and the half period is \SI{5}{\micro s}. It is known that the voltage across a PT is flipped between $\pm (V_S + 2V_D)$ every half vibration period, so the voltage flipping operation should finish in a time much less than the half period in order to avoid excessive wastage of generated energy. Fig. \ref{fig:1909_sshcwf} shows the waveforms of the $I_P$ and the voltage $V_{PT}$ during normal voltage flipping operations. The time used to flip the voltage $V_{PT}$ is labeled as $T_F$ and the half-period is $T/2$. Besides of the desired high voltage flip efficiency, we also want that the time $T_F$ is much less than $T/2$ in order to minimize the energy loss, shown as the black areas in the waveform of $I_P$. 

The voltage flip time $T_F$ depends on the number of capacitors employed. Assuming a k-stage SSHC is employed (with $k$ capacitors), there are $2k+1$ phases to flip the voltage as previously discussed. The time spent in each phase should be long enough for the charge to flow from one capacitor to the other. Assuming $C_P = \SI{100}{pF}$, hence $C_1 = C_2 = \dots = C_k = \SI{100}{pF}$. When the $C_P$ is connected to one of the capacitors, the time constant for the two-capacitor charging loop is given as $\tau = R_{ON} C_P /2$. To achieve full capacitor charging, the required time for each phase should be at least $5 \tau = \frac{5}{2}R_{ON} C_P$. As a result, the total voltage flip time for the $2k+1$ phases is $T_F = \frac{5}{2} R_{ON} C_P (2k+1)$. In order to achieve $T_F << T/2$, we assume $T_F = \frac{1}{10} \frac{T}{2}$. Hence the resulting equation is:

\begin{equation}
\frac{5}{2} R_{ON} C_P (2k+1) = \frac{1}{10} \frac{T}{2}
\end{equation}

We have $C_P = \SI{100}{pF}$ and $T = \SI{10}{\micro s}$. Assuming we employ an 8-stage SSHC rectifier with $k=8$, the required ON resistance of the two-capacitor path is calculated to be $R_{ON} = 117 \Omega$. According to Fig. \ref{fig:1909_hsscwkcaps}, $R_{ON}$ consists of the ON resistance of the two analogue switches, metal wires and vias (in CMOS process). In order to let the SSHC rectifier able to be used in the ultrasonic PT with $C_P = \SI{100}{pF}$ and with resonant frequency of \SI{100}{kHz}, the total ON resistance in the path should be designed to have $R_{ON} \leq 117 \Omega$. When working with PTs with larger $C_P$ or higher frequencies, the required $R_{ON}$ is smaller and more effort is needed for the switches, routing and via design.

\section{Conclusion} \label{1909_conclu}

This paper analyzes the SSHC rectifier for piezoelectric transducers (PT) used in ultrasonic wireless power transfer systems. In biomedical applications, such as wirelessly powered biomedical implants, both the system miniaturization and energy performance are the key considerations. Compared with the SSHI rectifier, the SSHC rectifier does not employ any inductor. Compared with the split-electrode (SE) SSHC rectifier, the conventional SSHC can be directly employed for ultrasonic PTs thanks to the small inherent capacitance of these sub-mm sized PTs. While the SE-SSHC requires thick oxide transistors in the CMOS process, the conventional SSHC can be designed in standard CMOS process. This paper also analyzes the design considerations regarding to PT capacitance, resonant frequency and stage number when employing the SSHC rectifier in piezoelectric ultrasonic WPT systems. The analysis presents the maximal ON resistance in the capacitor-to-capacitor charging loops to achieve expected performances.

\bibliographystyle{IEEEtran}
\def\baselinestretch{0.8}
\bibliography{./references}

\end{document}